# MODEL-INDEPENDENT PLOTTING OF THE COSMOLOGICAL SCALE FACTOR AS A FUNCTION OF LOOKBACK TIME

H. I. Ringermacher* and L. R. Mead*
Dept. of Physics and Astronomy, U. of Southern Mississippi, Hattiesburg, MS 39406, USA

**ABSTRACT**
In the present work we describe a model-independent method of developing a plot of scale factor $a(t)$ versus lookback time $t_L$ from the usual Hubble diagram of modulus data against redshift. This is the first plot of this type. We follow the model-independent methodology of Daly and Djorgovski (2004) used for their radio-galaxy data. Once the $a(t)$ data plot is completed, any model can be applied and will display accordingly as described in standard literature. We then compile an extensive data set to $z = 1.8$ by combining SNe Ia data from SNLS3 of Conley et al. (2011), High-z SNe data of Riess et al. (2004) and radio-galaxy data of Daly & Djorgovski (2004) to be used to validate the new plot. We first display these data on a standard Hubble diagram to confirm the best fit for ΛCDM cosmology and thus validate the joined data set. The scale factor plot is then developed from the data and the ΛCDM model is again displayed from a least-squares fit. The fit parameters are in agreement with the Hubble diagram fit confirming the validity of the new plot. Of special interest is the transition-time of the universe which in the scale factor plot will appear as an inflection point in the data set. Noise is more visible on this presentation which is particularly sensitive to inflection points of any model displayed on the plot unlike on a modulus-z diagram where there are no inflection points and the transition-z is not at all obvious by inspection. We obtain a lower limit of $z \geq 0.6$. It is evident from this presentation that there is a dearth of SNe data in the range, $z = 1-2$, exactly the range necessary to confirm a ΛCDM transition-z in the neighborhood of $z = 0.76$. We then compare a "Toy Model" wherein dark matter is represented as a perfect fluid with equation of state $p = -(1/3)\rho$ to demonstrate the plot sensitivity to model choice. Its density varies as $1/t^2$ and it enters the Friedmann equations as $\Omega_{dark}/t^2$ replacing only the $\Omega_{dark}/a^3$ term. The Toy Model is a close match to ΛCDM but separates from it on the scale factor plot for similar ΛCDM density parameters. It is described in an appendix. A more complete transition time analysis will be presented in a later paper.

* E-mail: ringerha@gmail.com and www.ringermacher.com: Lawrence.mead@usm.edu

## 1. INTRODUCTION

Traditionally, the Hubble diagram plots modulus against redshift, both of which are observational measurements. SNe Ia data are always seen this way. The chosen cosmological model is then fitted and secondary quantities such as the deceleration parameter, the transition redshift, the age of the universe, etc. are extracted by operations on the fitted parameters. The transition redshift of the universe is the redshift value at which the universe transitions from decelerating to accelerating. The fact that the universe is accelerating at all was discovered in 1998 and a Nobel Prize was awarded (Perlmutter, Schmidt & Riess 2011). The transition-z is thus a critical point that is highly model-dependent. Indeed, some authors (Lima et al. 2012) have even suggested that the transition-z be regarded as a new cosmological number. The location of this point is not obvious in a standard Hubble diagram because the distance modulus makes no noticeable changes at that location. In order to obtain the transition redshift one must evaluate the deceleration parameter at the point where it vanishes. Thus one must take second derivatives of noisy data - generally not desirable. Daly and Djorgovski (2003) comment on this as "a cardinal sin for any empirical scientist", but authors do it anyway.

An alternative approach is to utilize the Hubble diagram data to create a plot of the scale factor, $a(t)$, versus lookback time, $t_L$. This plot displays the inflection point at the transition time visually, unlike a modulus plot where the location of this point is unintuitive. Only one derivative need be taken on the scale factor plot to locate this point, thus reducing noise and permitting higher sensitivity to model discrimination. Scale factor plots are seen in every cosmology textbook but appear to be underutilized in the literature. The reason is apparently that it is assumed that a cosmological model must first be selected in order to calculate lookback time. In fact, that is not necessary. Daly and Djorgovski (2003, 2004) have developed a model-independent approach to calculate important cosmological parameters, for example the expansion parameter, $E(z)$ and the deceleration parameter, $q(z)$. They derive formulas for these, based upon estimates of the "dimensionless coordinate distances" of galaxies. We take this work a step further by analyzing lookback time similarly.

The primary purpose of this paper is to describe and demonstrate a model-independent approach to develop a scale factor-lookback time plot. This paper is organized as follows. We first present the theory for this approach demonstrating why a model is not needed, allowing one to plot empirical data. A red shift data set is then selected for the scale factor plot. In fact we combine SNLS3, 2011 SNe Ia data of Conley, et al. (2011) with the 2004 Radio Galaxy data of Daly and Djorgovski (2004) and some High-$z$ SNe Ia data of Riess, et al. (2004) to provide a baseline to $z = 1.8$. This data set is first validated on a standard Hubble diagram of modulus against red shift by displaying a least-squares fit of ΛCDM. The same data set is then converted for the scale factor versus lookback time plot. This process is described in detail. We then display the same ΛCDM model from a least-squares fit to the converted data to validate the $a(t)$ vs. $t$ approach. The two least-squares ΛCDM fits to the two types of plots must result in the same fitting density parameters in order to instill confidence in the model-independent approach – and they do. Finally, a "Toy Model" for dark matter is introduced and displayed on the same scale factor plot to demonstrate its sensitivity to model differentiation. The Toy Model is described in Appendix A. We will leave $a(t)$ data analysis to a later paper.

## 2. THEORY

We begin by writing the FRW metric for the ΛCDM model:

$$ds^2 = dt^2 - a(t)^2 \left( \frac{dr^2}{1-kr^2} + r^2 d\theta^2 + r^2 \sin^2\theta d\phi^2 \right) \quad (1)$$

We choose a flat 3-space from current measurements and set $k = 0$. We note that $r, \theta, \phi$ are "frozen" or comoving coordinates. However, they define a position for each galaxy observation imagined to span from the present to distant past thus representing a family of red shifts and coordinate distances with an implicit time dependence. A formal discussion of this point is presented below.

Lookback time is traditionally calculated from the following integral:

$$t_L = t_H \int_0^z \frac{dz'}{(1+z')E(z')}, \quad (2)$$

where
$$E(z) = \sqrt{\Omega_m(1+z)^3 + \Omega_k(1+z)^2 + \Omega_\Lambda} \quad (3)$$

is the Hubble parameter for ΛCDM and the density parameters are $\Omega_m$ for dark plus baryonic matter, $\Omega_k$ the curvature parameter and $\Omega_\Lambda$, the dark energy density parameter. In this paper we set $\Omega_k = 0$ for a flat universe. $t_H$ is the present Hubble time, $1/H_0$.

Let us examine this formula in detail. The scale factor is defined by $a(t) = 1/(1+z)$. Also we must have, by definition,

$$E(z) = \frac{\dot{a}(t)}{a(t)}, \quad (4)$$

where the overdot is the derivative with respect to light travel time (coordinate time), $t$. Clearly, associated with every observed red shift there must be a light travel time from that source. But from the above definitions alone it is clear that the integral (2) is simply

$$\int_{t_z}^{t_0} dt' = 1 - \frac{t_z}{t_H} = \tau_L, \quad (5)$$

where $t_z$ is the light travel time from the source at red shift z and $\tau_L$ is the dimensionless lookback time. Here we have normalized time with respect to the Hubble time so that the present time is $t_0 = 1$. From the metric, Eq.(1), the light travel interval along a fixed line-of-sight is:

$$dt = a(t)dr \quad (6)$$

This time interval is interpreted as the light travel time interval between two spatially consecutive SNe sightings of a family of observations. The space between the two observations expands such that the sum over all observations of z is the light travel time, $t_z$, from the most distant source to the nearby one at $a(t_0) = a(1) = 1$. One must also be certain that the intrinsic condition, $\dot{a}(1) = 1$, is also satisfied for a proper plot. It remains to describe the coordinate distance, r, in terms of time. We shall be working with several distance measures. Modulus, $\mu_0$, is a measure of luminosity distance, $D_L$ (Mpc) and is defined from:

$$\mu_0 = m - M = 5 Log\, D_L + 25 \tag{7}$$

The luminosity distance is defined from the comoving distance which is our metric coordinate distance, $r$:

$$D_L = r / a(t) \tag{8}$$

Thus,

$$dr = d[\, a(t)\, D_L\, ]$$

With distances normalized to the Hubble length, and time to Hubble time, our coordinate distance $r(t)$ is the same as the "dimensionless coordinate distance", $y(z)$, of Daly and Djorgovski and we may write, adopting their notation;

$$dy = d[\, a(t) \frac{D_L}{D_H}\, ], \tag{9}$$

where $D_H = c\, t_H$. We shall keep Eq. (9) in differential form because both $a(t)$ and $D_L$ vary with each SN measurement and we will analyze our data this way, consistent with Eq. (6). Finally, from Eqs. (5) - (9), we can write for the empirical dimensionless lookback time, $\tau_L$;

$$\tau_L = 1 - \int_{t_0}^{t_z} a(t)\, dy \tag{10}$$

together with $\qquad a(t) = 1/(1+z)$,

thus relating our plot to direct measurements of red shift and luminosity distance, a numerical procedure which will become clear in the $a(t)$ plot section. We next proceed to select a data set.

## 3. DATA SELECTION AND VALIDATION

In selecting data to validate our plots we desire as high a red shift range as possible. Since we base our approach on the work of Daly and Djorgovski, naturally we are strongly influenced by their work on radio galaxies as standard candles (Daly and Djorgovski 1994) We choose to combine 18 of their 20 RGs (excluding 3C405 and 3C427.1) out to $z = 1.8$ in Daly & Djorgovski (2004) with more recent 2011 SNLS3 SNe Ia data of Conley et al. (2011) which alone goes to $z = 1.4$. To help fill the sparse region between $z = 1$ and $z = 2$ we add High-z SNe Ia data of Riess, et al. (2004) for $z \geq 1$ also in Union 2.1 (Kowalski et al. 2008). Conley, et al. do not provide a determination of $H_0$ to permit scaling their data with regard to the estimated absolute magnitude of a Type Ia SN. Daly and Djorgovski do scale their 2004 data by combining it with the SNe data of Riess, et al. (2004). They use their own estimated Hubble constant of 66.4 km/s/Mpc from the Riess, et al. data. They obtain this value by examining the low-z ($z < 0.1$) linear Hubble diagram of Riess, et al. Thus their RGs are well-scaled to the Riess, et al. SNe.

We therefore choose to scale the Conley, et al. (2011) data with respect to the data of Daly and Djorgovski (2004) and the data of Riess, et al. (2004). The Conley, et al. data are considered very accurate with multiple corrections described in their work. We applied their corrections to obtain the corrected magnitude $m_{corr}$. We then compare the Conley, et al. and Riess et al. data to evaluate the supernova absolute magnitude noting that Daly and Djorgovski have successfully scaled to Riess, et al. We select 8 SNe in common to the two data sets (Table 1). We first estimate our own Hubble constant from the low-z Conley, et al. data. We included data out to $z = 0.1$ (133 points) for a low-z yet sufficiently large set for a high confidence Hubble fit. The present Hubble constant is found from the coordinate distance, $r$, given the luminosity distance, $D_L$, and red shift velocity, $v$, using the relation:

$$H_0 = \frac{v}{r} = \frac{cz}{a D_L} = \frac{cz(1+z)}{D_L} \tag{11}$$

We find $H_0 = 69.0$ km/s/Mpc for the Conley, et al. set. This is only for our purposes in scaling calculations and is not meant to fix their scale. Conley (2012) pointed out that care must be taken when making such estimates due to the extreme sensitivity of the data to the choice of the SN Ia magnitude, $M_0$. We later rescale the Conley, et al. data by normalizing the coordinate distances to $H_0 = 66.4$ for consistency. We then compared the Conley, et al. and the Riess, et al. data for the 8 SNe selected and estimated the least squares SN Ia absolute magnitude, $M_0$, necessary to give the two sets identical moduli for those points. Averaging the 8 values, we find $M_0 = -19.19 \pm 0.13$ in agreement with the Riess, et al. estimate of $M_0 = -19.3$. This absolute magnitude is then subtracted from $m_{corr}$ to obtain the modulus for the combined data. A slight correction to $M_0 = -19.24$, within our error, was made to best fit our combined SNe to high z. The joined three sets are shown in Fig. (1). Also shown in Fig. (1) is the traditional fit of ΛCDM with least squares density parameters, $\Omega_\Lambda = 0.728$ and $\Omega_m = 0.272$, essentially the WMAP values, thus confirming the quality of the data set. The combined Modulus-z data set is available (CombinedDat2013).

**Table 1: 8 SNe in common with the Riess, et al. and Conley, et al. data**

| source | Riess $\mu_0$, modulus | Conley $m_{corr}$ | $M_0$ |
|---|---|---|---|
| sn1999cc | 35.73 | 16.48 | -19.26 |
| sn1999gp | 35.36 | 16.20 | -19.15 |
| sn2000ca | 35.96 | 15.92 | -19.03 |
| sn2000cf | 36.11 | 17.01 | -19.11 |
| sn2000cη | 34.03 | 14.61 | -19.43 |
| sn2001ba | 35.58 | 16.56 | -19.03 |
| sn2001cn | 34.02 | 14.75 | -19.28 |
| sn2001cz | 34.09 | 14.87 | -19.23 |

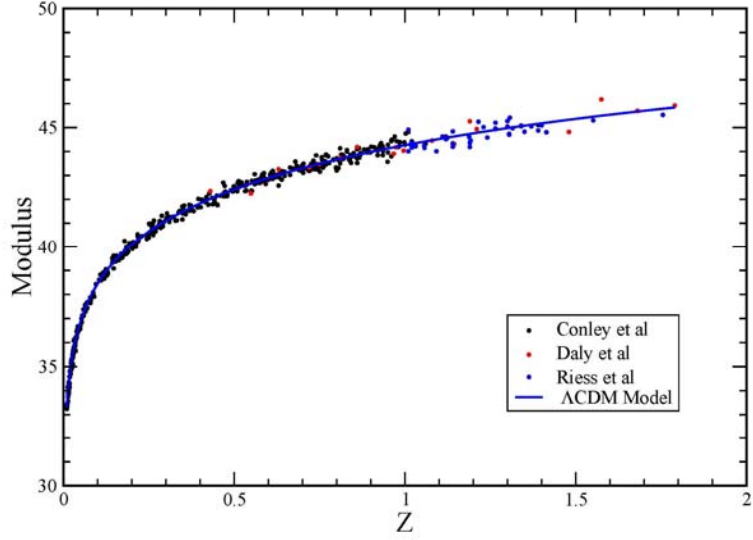

**Fig. 1.** Joined SNe Ia data sets of Conley, et al., Daly and Djorgovski and Riess, et al. with least squares fit of ΛCDM. Modulus is from Eq. (7).

## 4. PLOTTING THE SCALE FACTOR AGAINST LOOKBACK TIME

We first calculate the lookback time. We will follow Eq. (7-10) very closely and will present a table showing a sample calculation. We assume we have the redshift, $z$, and the luminosity distance in Mpc, $D_L$. $D_L$ is calculated from Eq. (7), given typical modulus data, $\mu_0$. Table 2 shows a series of measurements sorted by ascending $z$ in column 1. Shown are a set starting with the lowest z values followed by a gap jumping to around z = 1 in order to show the changes in the running sum over column 5 to get the lookback time in column 6. The labeled columns are calculated as follows:

Column 1:    $z$ given
Column 2:    $a = 1/(1+z)$
Column 3:    $D_L$ in Mpc from Eq.(7) given modulus $\mu_0$
Column 4:    $Y = a \dfrac{D_L}{D_H}$,   $D_H = c/66.4 = 4514.94\,\text{Mpc}$   ($H_0 = 66.4\,\text{km/s/Mpc}$)
Column 5:    $a \cdot delta Y_i = a_i \cdot (Y_i - Y_{i-1})$
Column 6:    $LookbackT_j = 1 - \sum_{i=1}^{j} a_i \cdot (Y_i - Y_{i-1})$
Column 7:    $LookbackTcorr_j = LookbackT_j - 0.010061$

**Table 2: Sample calculation of lookback time**

| z | a | $D_L$, Mpc | Y | a*deltaY | *LookbackT* | *LookbackTcorr* |
|---|---|---|---|---|---|---|
| 0.01006 | 0.9900402 | 44.59782 | 0.009779452 | | | |
| 0.01029 | 0.98981481 | 45.17934 | 0.009904711 | 0.000124 | 0.999876016 | 0.989815016 |
| 0.01055 | 0.98956014 | 46.90968 | 0.010281411 | 0.000373 | 0.99950325 | 0.98944225 |
| 0.0109 | 0.98921753 | 48.76659 | 0.010684696 | 0.000399 | 0.999104313 | 0.989043313 |
| 0.01113 | 0.98899251 | 49.50145 | 0.010843237 | 0.000157 | 0.998947517 | 0.988886517 |
| 0.01231 | 0.98783969 | 67.30763 | 0.014726474 | 0.003836 | 0.995111501 | 0.985050501 |
| 0.01334 | 0.98683561 | 55.71892 | 0.012178549 | -0.00251 | 0.997625885 | 0.987564885 |
| 0.01354 | 0.98664088 | 51.35715 | 0.011222976 | -0.00094 | 0.998568692 | 0.988507692 |
| 0.01366 | 0.98652408 | 61.35487 | 0.013406172 | 0.002154 | 0.996414917 | 0.986353917 |
| 0.0138 | 0.98638785 | 61.96632 | 0.013537903 | 0.00013 | 0.996284978 | 0.986223978 |
| * | * | * | * | * | * | * |
| 1 | 0.5 | 7175.263 | 0.794613375 | -0.01996 | 0.442087038 | 0.432026038 |
| 1.002 | 0.4995005 | 8922.165 | 0.987084176 | 0.096139 | 0.345947777 | 0.335886777 |
| 1.01 | 0.49751244 | 9196.367 | 1.013370481 | 0.013078 | 0.332870013 | 0.322809013 |
| 1.02 | 0.4950495 | 7673.713 | 0.841399368 | -0.08513 | 0.418004227 | 0.407943227 |
| 1.02 | 0.4950495 | 6891.824 | 0.755667683 | -0.04244 | 0.460445656 | 0.450384656 |
| 1.031 | 0.49236829 | 6971.197 | 0.760230787 | 0.002247 | 0.458198928 | 0.448137928 |
| 1.06 | 0.48543689 | 7307.563 | 0.785693913 | 0.012361 | 0.445838187 | 0.435777187 |

There are several points to note in order to properly calculate lookback time. Column 5 clearly shows the presence of noise. This is effectively smoothed by the integration in column 6 but lookback time nevertheless carries the noise. More importantly, two criteria must be satisfied: [A]; $a(1)=1$ and [B]; $\dot{a}(1)=1$. An inspection of the table at row 2 shows that $a \neq LookbackT$. That is because this a is not the one for the present time, but rather for the nearest measured z. So there is an apparent time gap of $0.010061$. This is subtracted from *LookbackT* to generate *LookbackTcorr* in column 7. In effect this gap is an amount $\Delta a(t) \simeq z$ by virtue of the definition of $a(t)$ and is considered an integration constant. Condition [A] is then satisfied and the data is centered on the present time. *LookbackTcorr* is used in the final plot but will be referred to as lookback time. The slope at time 1 is determined by adjustment of the Hubble constant. In the present case the slope is approximately 0.98 and thus satisfies [B]. For the final plot, Table 2 is sorted again by the corrected lookback time. Thus the random noise present in the lookback time is transformed to scatter in $a(t)$.

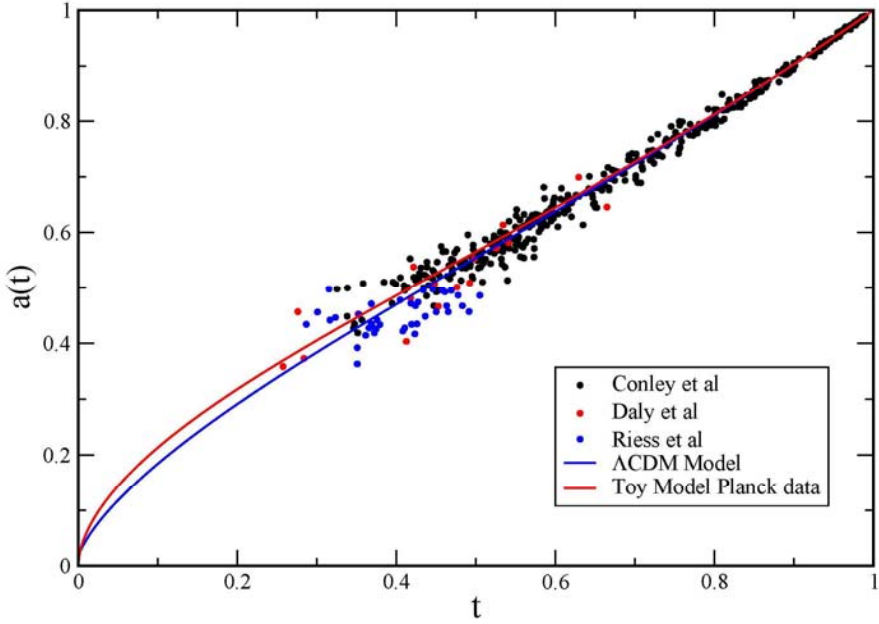

**Fig. 2.** Plot of scale factor against lookback time for the combined data set. The blue curve is the least-squares fit for ΛCDM. The red curve is the "Toy Model" for Planck density parameters.

Figure 2 shows the combined data set plotted as scale factor against corrected lookback time. Also shown on the plot is the least-squares fit of ΛCDM with resulting density parameters: $\Omega_\Lambda = 0.735$ and $\Omega_m = 0.265$. These values are extremely close to those from the Modulus plot, Fig.1, thus supporting the validity of the lookback time calculation. The R-squared goodness of fit for ΛCDM in Fig.2 is 0.98. Also shown on the plot is a "Toy Model" (Appendix A) wherein dark matter is represented as a perfect fluid with equation of state $p = -1/3\,\rho$. Its density varies as $constant/t^2$ and it enters the Friedmann equations as $\Omega_{dark}/t^2$ replacing only the $\Omega_{dark}/a^3$ term in ΛCDM. Otherwise it is calculated in exactly the same way as the ΛCDM scale factor. With this replacement, a new solution for the Toy $a(t)$ can be found. The Toy model is not a least-squares fit in order to demonstrate the separation on the plot. The Planck parameters were used for the Toy Model: $\Omega_\Lambda = 0.68$, $\Omega_m = 0.05$ and $\Omega_{dark} = 0.27$. The Toy Model is a close match to ΛCDM. A Toy least-squares fit would be indistinguishable from ΛCDM and the fit parameters would be: $\Omega_\Lambda = 0.61$, $\Omega_m = 0.05$ and $\Omega_{dark} = 0.34$. Both curves lie well within the data scatter for their current parameters.

## 5. TRANSITION-TIME TO AN ACCELERATING UNIVERSE

A full analysis of the $a(t)$ data will be presented in a later paper. However, a simple inspection of the data suggests the $a(t)$ inflection point, or transition-time, lies conservatively at $t \leq 0.6$ ($z_t \geq 0.57$). At later times the slope is accelerating. A simple quadratic least-squares data fit matches the two closely spaced model curvatures. At earlier times the inflection region is very broad and the data must eventually turn over toward the origin. The ΛCDM transition-time for $\Omega_\Lambda = 0.735$ and $\Omega_m = 0.265$ is expected at $t \simeq 0.514$ corresponding to $z_t = 0.77$. Riess, et al. have stated a value of $z_t = 0.46 \pm 0.13$ (2004) and $z_t = 0.426^{+0.27}_{-0.089}$ (2007). Daly and Djorgovski have independently found $z_t \approx 0.45$ (2004) and, with an expanded data set, $z_t = 0.78^{+0.08}_{-0.27}$ (2008). Lima, et al. (2008) also checked the Riess, et al. (2004) data and confirmed their estimate within error. Cunha and Lima (2008) examined Astier, et al. (2006) SNLS data and found $z_t = 0.61$. In the same paper they also examined the data of Davis, et al. (2007) and found $z_t = 0.60$. They separately examined Union data (Kowalski, et al. 2008) and found $z_t = 0.49^{+0.14}_{-0.07}$. Transition times tend to be clustered around $z_t = 0.45$ and $z_t = 0.60$. The Daly and Djorgovski (2008) value, $z_t = 0.78^{+0.08}_{-0.27}$, agrees with ΛCDM but has extremely wide error bars. The wide variation in transition times would indicate a problem, or as Lima, et al. (2008) have put it, "this could be seen to raise some mild flags with the standard ΛCDM model". Clearly the data is noisy and simply insufficient to determine this number precisely at the present time. More data in the range $1 < z < 2$ would be helpful.

## 6. CONCLUSIONS

We describe a novel model-independent approach to plot the cosmological scale factor against lookback time. This is a new way of plotting empirical standard candle data as opposed to the usual Hubble diagram. We selected and joined two SNe data sets together with Radio-Galaxy observations to create a standard candle baseline to $z = 1.8$ to be utilized in validating the new plot. The data was first plotted in the usual form of modulus against red shift and the ΛCDM model was seen to present a classic fit through the data, thus validating the joined data set. The $a(t)$ plot was then constructed and the same ΛCDM model was found to fit well, thus validating the new plot. A "Toy Model" was also constructed and superposed on the scale factor plot using Planck parameters to compare against ΛCDM. The match was surprisingly good – well within the plot scatter but the new plot successfully discriminated the subtle difference. It is clear from inspection of the $a(t)$ plot that there is a dearth of data between $z = 1$ and $z = 2$ thus resulting in a wide range of estimates of the transition-z and apparently spanning the entire range of $z_t = 0.45$ to $z_t = 0.78$ biased in general toward the lower values. This may simply be noisy data or it might suggest tension with the ΛCDM model.

# APPENDIX A

## TOY MODEL FOR DARK MATTER

E. Kolb, in 1989, describes a "coasting universe" with a dominant form of matter that he refers to as "K-matter" (Kolb, 1989). K-matter derives from the Friedmann equation for the FRW metric, Eq.(1), for a matter density that varies as $1/a(t)^2$. Any density of this form, in effect, enters the Friedmann equation as a simple constant curvature contribution, $k$ – hence his name, "K-matter". Kolb found that the equation of state of this fluid is $p = -1/3\,\rho$ with the result that the scale factor acceleration vanishes since we have:

$$\frac{\ddot{a}(t)}{a(t)} = \frac{4\pi G}{3}(\rho + 3p) = 0 \tag{A1}$$

He goes on to describe various universes dependent upon the curvature and properties of those universes such as effects on red shift, etc. The Concordance model did not exist. Today, a "coasting" universe model based on "K-matter" has been rejected since the curvature of the universe has been measured as flat ($k = 0$).

However, we now take a somewhat different view. We know that the FRW spatial curvature vanishes with high confidence. K-matter in the form described by Kolb would add nothing to this scenario. However, we might consider a form of K-matter as a new type of time-dependent matter, replacing dark matter. This is essentially "coasting dark matter". Kolb does not discuss this direct consequence. The appropriate matter density is inserted into the Friedmann equation for metric (1) as $\Omega_{dark}/t^2$ replacing $\Omega_{dark}/a^3$ in ΛCDM. The baryonic matter and dark energy are left intact. For this new density, we solve the Friedmann equation numerically, valid for all times. This new solution remains consistent with the constraint $k = 0$; $\Omega_k = 0$ over all times as found observationally. We call this alternative model our "Toy Model". A least squares fit of our Toy Model makes it indistinguishable from ΛCDM within the width of the plot line using density parameters; $\Omega_\Lambda = 0.61$, $\Omega_b = 0.05$ and $\Omega_{dark} = 0.34$. Figure 2 simply displays the Toy Model as a contrasting model for a choice of Planck density parameters. Further pursuit of this model is outside the scope of our paper.

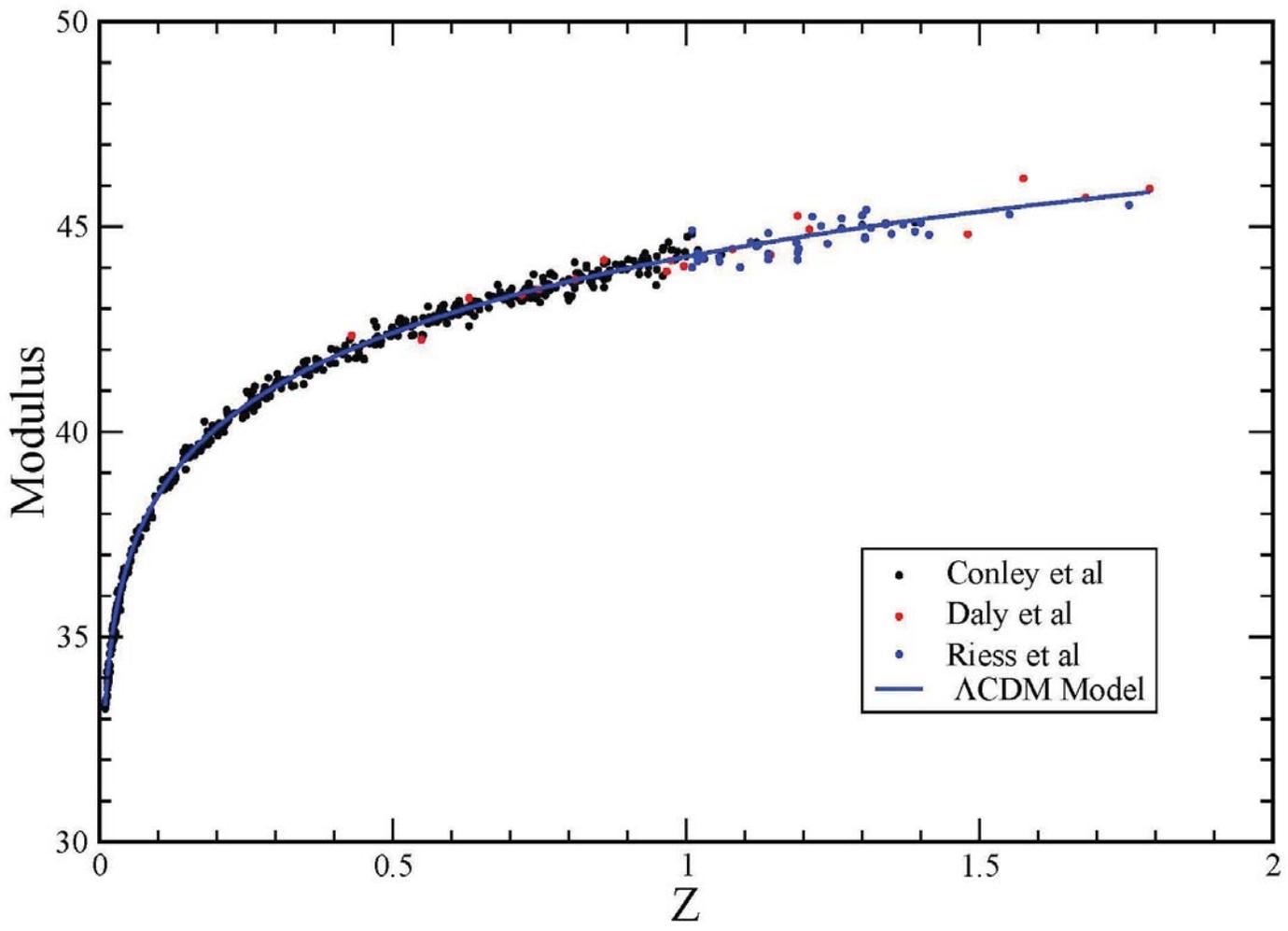

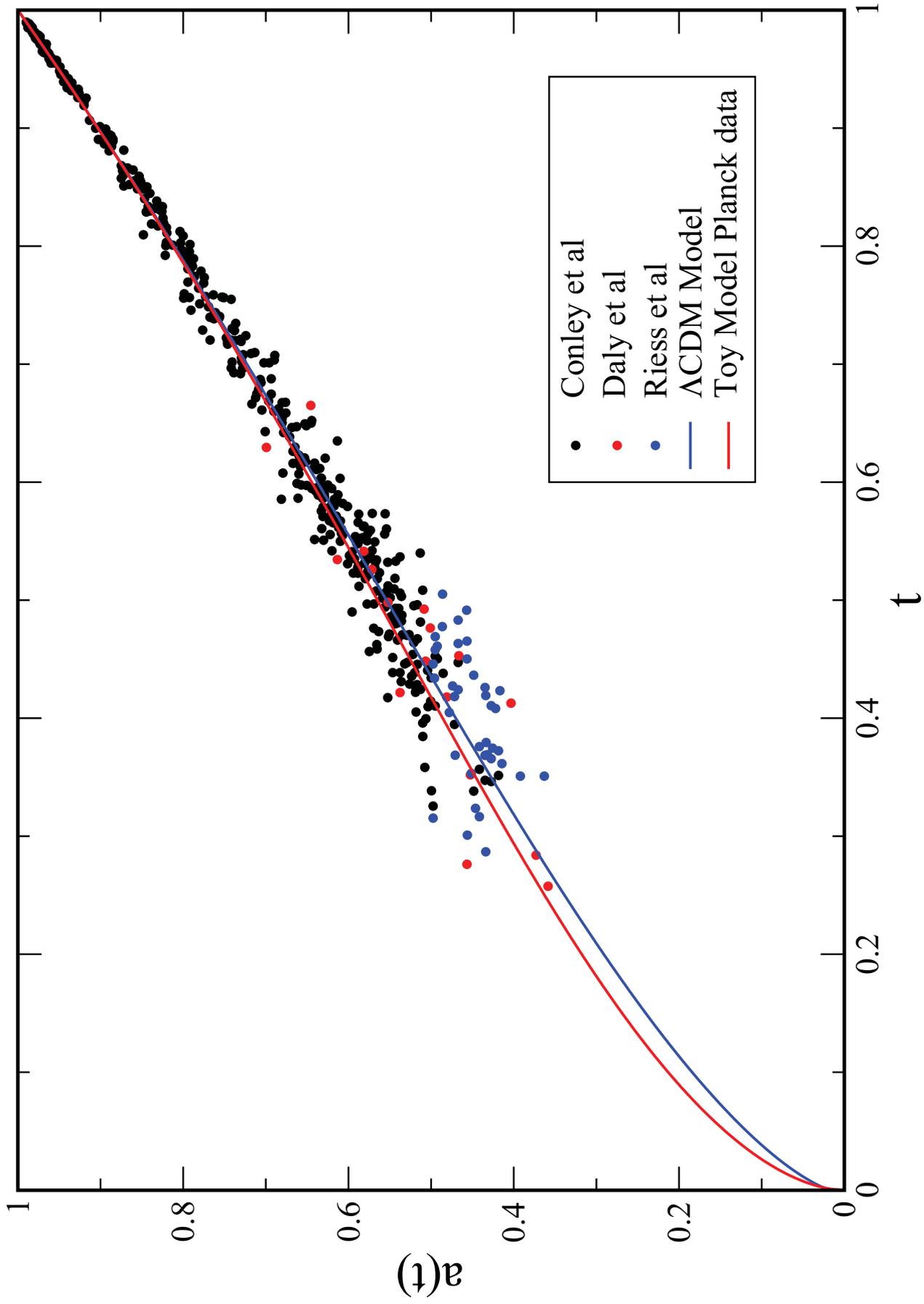